 \documentclass[final,5p,times,compress,twocolumn]{elsarticle}

\usepackage{amssymb}
\usepackage{amsmath}
\usepackage{lipsum}
\usepackage{hyperref}
\usepackage{color}
\usepackage{soul}
\usepackage{widetext}

\journal{Physics Letters B}

\date{\today}
\begin{document}

\begin{frontmatter}



\title{Massive hybrid stars within the extended three-flavor quark-meson diquark model}


\date{\today}

\author[a]{Jens O. Andersen}
\ead{jens.andersen@ntnu.no}

\author[a,b]{Manuel Linares}

\ead{manuel.linares@ntnu.no}

\author[a]{Mathias P. Nødtvedt}

\ead{mathias.p.nodtvedt@ntnu.no}
            
\affiliation[a]{organization={Department of Physics, Faculty of Natural Sciences, Norwegian University of Science and Technology},
            addressline={H{\o}gskoleringen 5}, 
            city={Trondheim},
            postcode={N-7491}, 
country={Norway}}

\affiliation[b]{organization={ Departament de F{\'i}sica, EEBE, Universitat Polit{\`e}cnica de Cataluny},
            addressline={Av. Eduard Maristany 16}, 
            city={Barcelona},
            postcode={E-08019}, 
country={Spain}}

\begin{abstract}
We discuss the properties of the extended three-flavor quark-meson diquark (EQMD) model as a renormalizable low-energy effective model for QCD. The effective
degrees of freedom are quarks, scalar- and pseudoscalar mesons, 
diquarks, vector- and axial-vector mesons. We calculate the equation of state (EoS) in the mean-field approximation at $T=0$ imposing charge neutrality for electric and color charges. We match the EoS with a low-density nuclear equation of state. We discuss how the choice of parameters in the model affects the EoS and thereby the mass-radius for hybrid stars.
We show that it is possible to construct hybrid stars whose masses and radii are in agreement with recent astrophysical observations
and perturbative QCD (pQCD).
The addition of vector and axial vector mesons to the quark-meson diquark is essential, since it makes the EoS sufficiently stiff for intermediate densities.
Our results show that stars with a mass larger than $M\sim2M_{\odot}$ can have a quark core with a central density $n_B\geq 3.9n_{\rm sat}$,
where $n_{\rm sat}\approx0.165$fm$^{-3}$ is the saturation density.
The speed of sound has a double-peak structure and relaxes to the conformal limit from above for large baryon chemical potentials $\mu_B$. This structure is caused by the decrease in the mass of the $s$ quark as $\mu_B$ increases.

\end{abstract}




\end{frontmatter}




\section{Introduction}
\label{introduction}
QCD in extreme conditions has received a lot of attention for decades.
In the case of dense QCD, this is largely spurred by the interest in understanding neutron stars. Neutron stars
are among the densest objects in the universe, with a maximum mass of $M\simeq2.3~M_{\odot}$ and a corresponding radius of $R\simeq12$ km.
Within general relativity, the mass and radius of a star are found
by solving the Tolman-Oppenheimer-Volkov (TOV) equation. This requires knowledge
of the equation of state. Masses and radii are not the only interesting quantities in neutron star physics. Understanding the tidal deformabilities of coalescing binary neutron stars is required
in order to model the gravitational waves (GW) they emit. Moreover, the ringdown phase has also been explored as a possible probe for the neutron star EoS at even higher densities than what can be constrained through the inspiral phase \cite{EckerNature}.
Neutron stars are the only objects where one can 
obtain indirect access to the equation of state for strongly interacting matter at high density and low temperature.
It is expected that hadronic matter undergoes a
transition from nuclear degrees of freedom to quark degrees of freedom as the density increases. An interesting
question is whether there are one or more deconfined phases in the most massive neutron stars that have been observed. 

Although QCD is established as the fundamental theory of strong interactions,
finite baryon density calculations are hampered by the infamous sign problem. 
The Dirac determinant is complex for nonzero values of $\mu_B$.
This implies that lattice simulations using standard importance sampling cannot be used, and one must use other methods to calculate the thermodynamic quantities. One first-principle method that can be used is perturbation theory at large density. Perturbation theory at large baryon density
has a long story that goes back to the seminal papers in the late 1970s,
where the thermodynamic quantities to ${\cal O}(\alpha_s^2)$ for massless
QCD were calculated~\cite{dense1,dense2,dense3,dense4}.
The strong coupling constant $\alpha_s$ is small at large densities, so that
the perturbative series is expected to converge 
for densities of at least 25-40~$n_{\rm sat}$.
In recent years, significant progress has been made in calculating the thermodynamical properties. 
Finite quark-mass effects have been
included in~\cite{finitemass1,finitemass2}.
Currently, thermodynamic quantities are
known to order $\alpha_s^3\log{\alpha_s}$~\cite{alpha3} and an outline of
the full $\alpha_s^3$ calculation was recently given in~\cite{vuorinen}.
\indent
In the weak-coupling calculations above, the expansion is about the trivial ground state. However, the true ground state of high-density QCD is not characterized by vanishing condensates, rather it is characterized by nonzero color superconducting gaps~\cite{bailin}: there is an attractive channel in one-gluon exchange rendering the Fermi surface unstable giving rise to the formation of Cooper pairs and pairing gaps. 
The phase diagram at high density (and low temperature) turns out to be very rich. Depending on the density and the details of the models, the phases are normal quark matter, 2SC phase, color-flavor locked (CFL) phase, and even inhomogeneous phases. The ideal CFL phase, where all three flavors and colors form pairs with a common gap, is known to be the correct ground state at very high densities.
For a detailed discussion, see, e. g.~\cite{alfordrev,hatsuda}.
The gaps are estimated to be 50-150 MeV for densities relevant for neutron stars~\cite{alfordrev}. A bound on the gap based on
astrophysical constraints was first derived in~\cite{rajakurk};
based on "reasonable assumptions" (such as $c_s^2\leq{1\over2}$ at the center of the most massive stars) and incorporating the leading-order
CFL correction to the N$^2$LO pQCD pressure,
they obtain an upper bound at $\mu=2.6$ GeV of 216 MeV~\cite{rajakurk}.
Including the NLO corrections of the CFL gap, which turn out to be
large, the upper bound has been estimated to be $140$ MeV~\cite{geisel2}.

The other extreme, namely that of low density, can also be described in a
model-independent way. Chiral effective field theory 
($\chi$EFT)~\cite{weinberg} is the correct low-energy theory for QCD that generalizes chiral perturbation theory 
($\chi$PT)~\cite{gasser} to include nuclear degrees of freedom.
It provides a systematic low-energy expansion for nuclear interactions.
The current state-of the-art calculations
give a reliable description for baryon densities up to or slightly above the nuclear saturation density $n_{\rm sat}$~\cite{ceft,ceft3,ceft1,ceft2,ceft4}. Since the density in the center
of the most massive neutron stars is expected to be 
$\approx5-8~n_{\rm sat}$, this leaves us with a large range of densities that must be described by other methods.

In the hadronic phase, there are a large number of nuclear models that describe nuclear matter at medium densities, for example, relativistic mean field theories and relativistic density functional models. For a review, see~\cite{revnuclear}.
The deconfined phase is often described by an extension of the MIT bag model or by the Nambu-Jona-Lasinio (NJL) model. The latter is a (perturbatively) non-renormalizable low-energy model for QCD with quark degrees of freedom. Its ultraviolet divergences are typically regulated using a sharp cutoff or a form factor.
The conventional way of regularizing the NJL model leads to regularization artifacts at finite density, for example, the diquark
gaps vanish for $\mu=600-700$ MeV, depending on details.
These issues have largely been resolved using a 
medium-separation scheme~\cite{mss} or by implementing
renormalization-group ideas~\cite{njl}.

There has also been enormous progress on neutron star research from the observational side in recent years.
Several groups have presented evidence for massive neutron stars reaching $2-2.3~M_\odot$ \cite{Demorest_nature,Antoniadis_Science,Linares_ApJ,Romani_ApJ}, based on radio and optical observations of millisecond pulsars in binary systems. 
Modeling the observed X-ray pulse profiles with thermal ``hot spots" on the surface
can provide simultaneous constraints on the masses and radii of neutron stars \cite{Watts19}. 
At least two systems have been studied in detail using this technique and
the Neutron Star Interior Composition Explorer (NICER)~\cite{Gendreau16}.
Similarly, the breakthrough in gravitational-wave (GW) astronomy over the last decade has opened up a new avenue to probe
the properties of dense matter. For example, the gravitational-wave signal from the latter stages of the  merger of two
neutron stars contains information about the tidal deformability $\Lambda$ \cite{gw}. This parameter again depends on the EoS of the two stars.

In this paper, we will use the extended three-flavor quark-meson diquark (EQMD) model to describe the deconfined phases of quark matter. The quark-meson diquark (QMD) model was first introduced in two-color QCD~\cite{lorenz}. It has mesons, quarks, and diquarks as effective
degrees of freedom using mean-field and functional renormalization group (FRG) methods. The QMD model was also discussed in some
detail in~\cite{braun1,braun2} and further investigated in~\cite{braun3}
in the context of the FRG.
In Refs.~\cite{us,us2,us3}, we studied the model
in the mean-field approximation, including only renormalizable interactions (see also~\cite{myra}).
The model has a number of coupling constants. The parameters in the mesonic sector are determined by matching model predictions to the observed meson masses and decay constants using the on-shell renormalization scheme. 
To be consistent, the matching must be done in the same approximation as the calculation of the thermodynamic potential. In the present case, we include quark loops, but treat mesons and diquarks at tree level.
The remaining parameters of the Lagrangian can be varied. By judiciously choosing the parameters, physically sound results were obtained~\cite{us,us2,us3}.
In the present paper, we extend the model by including a minimal vector
meson channel extension. These
new degrees of freedom are essential for constructing an EoS that is stiff enough at intermediate densities to support the most massive neutron/hybrid stars. We have then matched the high-density equation of state to a low-energy nuclear equation of state. The resulting EoS is used to calculate the speed of sound squared and the mass-radius relation for hybrid stars.

\section{Three-flavor extended quark-meson diquark model}
We now briefly discuss the most important aspects of the quark-meson diquark model and its extension to include vector and axial vector mesons. For a detailed discussion of the QMD model, we refer to~\cite{us3}. As mentioned in the introduction, the low-energy effective degrees of freedom are mesons, quarks, and diquarks.
In three flavor massless QCD, the symmetry group of the Lagrangian 
is $SU(3)_L\times SU(3)_R\times SU(3)_c\times U(1)_B\times U(1)_A$. For finite quark masses, the $SU(3)_L\times SU(3)_R$
chiral group is reduced to some subgroup that depends on the quark
masses. In the color superconducting (CSC) phases, the symmetry-breaking
pattern involves the local gauge symmetry.~\footnote{According to Elitzur's theorem, gauge symmetries cannot be broken spontaneously~\cite{elitzur}, however this wording is standard.}
For example, in the 
2SC phase, the breaking is $SU(3)_c\rightarrow SU(2)_c$ giving rise to 
five massive gluons via the Higgs mechanism.
In effective models, such as the NJL model and the quark-meson diquark model, there are no gluons and all the symmetries are global. The symmetry-breaking pattern in the CSC phases is the same as in QCD, however, since the symmetries are global, this leads to the
appearance of Goldstone bosons (GB) that are not present in the real world.
For example, in the 2SC phase there are three massless excitations, one with a linear dispersion relation (type A) and two with a quadratic dispersion relation (type B)~\cite{us2}. This result is in accordance with general counting rules for type A and 
type B Goldstone bosons established
over the past decade~\cite{gb1,gb2,gb3,gb4}.

The mesonic degrees of freedom are collected in a $3\times3$
matrix $\Sigma$. The matrix includes $\sigma$, the three pions, the
four kaons, $\eta$, and their parity partners $a$, $\kappa$, and 
$\eta^{\prime}$. The left-handed and right-handed diquark degrees of freedom are organized in $3\times3$ matrices as well, 
denoted by $\Delta_{L,R}$, where the rows correspond to color and 
columns to flavor. The quark fields are $\psi_a^i$, where $a$ is a color index, and $i$ is a flavor index. The new axial and vector mesonic degrees of freedom are also organized in two $3\times3$ matrices, denoted by $L^{\mu}$ and $R^{\mu}$, respectively. 
The terms allowed in the Lagrangian in the (extended) QMD Lagrangian are dictated by the symmetries and the requirement of perturbative renormalizability. 
The Lagrangian with axial and vector mesons, but without diquarks was 
discussed in detail in \cite{Rischke_LinearSigma_vector}.
Since the Lagrangian has a large number of new terms, we decided
to extend the Lagrangian minimally by adding only mass and quartic terms
for the new degrees of freedom.
The transformation rules for the different fields are
\begin{eqnarray}
\Sigma & \rightarrow& U_L\Sigma U_R^\dagger\;,\\
\psi_{L,R}&\rightarrow&U_{L,R}U_c\psi_{L,R}\;,\\
\Delta_L &\rightarrow& U_L\Delta_LU_c^T\;,\\
\Delta_R &\rightarrow &U_R\Delta_RU_c^T\;,\\
L^{\mu} &\rightarrow &U_LL^{\mu}U_L^{\dagger}
\;,\\
R^\mu&\rightarrow& U_RR^\mu U_R^\dagger\;, 
\end{eqnarray}
where $U_L$, $U_R$, and $U_c$ are left-handed, right-handed, and color transformations, respectively. 

Using these transformation properties and the cyclicity of the trace of a matrix, we can write down the invariant terms. For example, powers of $\langle\Sigma^{\dagger}\Sigma\rangle$ and 
powers of $\langle\Delta_{L,R}^{\dagger}\Delta_{L,R}\rangle$, where the symbol $\langle A\rangle$ means the trace of the matrix $A$, are invariant. 
There is a mass parameter $m^2$ for the mesons and two quartic couplings $\lambda_1$ and $\lambda_2$.
There is a mass parameter $m_{\Delta}^2$ for the diquarks and four
different quartic coupling constants
denoted by $\lambda^{\Delta}_i$ with $i=1,2,3,4$. There are three
quartic scalar-diquark couplings denoted by $\lambda_i$ with $i=3,4,5$. The quark-diquark coupling is denoted by $g_{\Delta}$.
The parameters in the effective Lagrangian are running masses and couplings. Using the solutions to the renormalization group equations, we obtain a mean-field expression for the renormalized thermodynamic potential
that is independent of the renormalization scale. The complete potential at \(T=0\) and can be found in~\cite{us3}. To extend the three-flavor QMD model, we have included three main terms in our Lagrangian. This leads to three changes in the final thermodynamical potential. We have a mass parameter \(m_\omega^2\) and a quartic coupling \(\lambda_\omega\) that multiply the appropriate powers of the two vector meson
condensates $\omega_l$ and $\omega_s$.
In addition we have coupled the new degrees of freedom to the quarks through the coupling \(g_\omega\) which effectively leads to a shift in the chemical potentials of the quarks \(\mu_f\rightarrow \mu_f - g_\omega\omega_f\). See \cite{ParityDoublet_Recci} for a recent incorporation in the parity-doublet model which is similar to the quark-meson model,  and \cite{golimi} for an incorporation in a two-flavor quark-meson model. 

In the Lagrangian, there are a total of nine chemical potentials, one for each color and flavor, denoted by $\mu_{ia}$. However, they are not all independent, but can be expressed in terms of 
the chemical potentials $\mu_B$, $\mu_e$, 
$\mu_3$, and $\mu_8$, which correspond to the conserved baryon and electric
charges $Q_B$, $Q_e$ and the two color charges $Q_3$ and $Q_8$.
Local charge neutrality is imposed by hand by requiring ${\partial\Omega\over\partial\mu_X}=0$ for $X=e,3, 8$ with $\Omega$ being the thermodynamic potential. Since there is a background of electrons,
we add their contribution $\Omega_1^e=-{4\mu_e^4\over3(4\pi)^2}$ 
(we take the electrons to be massless) to the 
thermodynamic potential.
The four chemical potentials, together with the three neutrality conditions, leave us with a single independent chemical potential $\mu_B$. We often
refer to the quark chemical potential $\mu$ which is equal to ${1\over3}\mu_B$.
Finally, note that charge neutrality is imposed by hand due to the lack of gauge fields in the model. In contrast, charge neutrality in QCD is obtained dynamically~\cite{alfordrev}. The zeroth component of the gauge fields acts as chemical potentials, and their nonzero expectation values are driven to values such that the charges are zero.
\vspace{-1mm}
\section{Construction of the equation of state}
To study hybrid stars, we need a realistic nuclear equation of state at both low and high densities. CompOSE~\cite{compose} is an online repository of EoSs
for use in nuclear physics and astrophysics, both in the hadronic and deconfined phases. 
We will briefly discuss our choice of nuclear EoS in the
next section.

There are several ways of constructing an EoS for compact stars.
The Maxwell construction is the most commonly used~\cite{glendenning}, which is relevant for one-component systems. In this approach, one connects a low density EoS to a high density EoS when the pressures are equal, which may result in a kink for the total pressure and a jump in the energy density.
In one-component systems, charge neutrality is imposed locally.
If we have mixed phases, a Gibbs construction is 
appropriate~\cite{glendenning}. In this case, the fraction of each phase
depends on the chemical potential and typically has a nonzero charge.
Therefore, charge neutrality is imposed globally.
Finally, we briefly mention a third possibility, namely interpolation~\cite{baymrev}. In this approach, we trust the nuclear equation up to a certain number density $n_{\rm low}\sim2n_0$. Similarly, we trust the equation of state for the deconfined phase down to a certain number density $n_{\rm high}\sim4$--$7n_0$. In the region between these two densities, we connect the pressure using interpolation.
In this picture of quark-hadron continuity, the hadronic  matter smoothly turns into quark matter as density increases.
The result is a so-called unified equation of state.
If the interpolation is smooth, the density is continuous, if the
pressure has a kink, the transition is first order.

Phenomenologically, we are free to choose a bag constant \(B\). A low bag constant leads to a stiffer equation of state, 
which can support
a more massive hybrid star, while a higher bag constant leads to a softer equation of state and therefore a less massive neutron star. 
However, there is a limit to how low a bag constant we can choose. If our choice of bag constant is too low \(P_{\rm nuc} \leq P_{\rm EQMD}\) for all \(\mu\) and no Maxwell construction is possible. 
Instead of focusing on the pressure, we focus on the number density.
The chemical potential \(\mu_t\) for switching from the nuclear EoS to the quark EoS is determined by demanding their number densities
to be equal
\begin{eqnarray}
\label{conti}
    n_{\rm EQMD} (\mu_t)= n_{\rm nucl}(\mu_t)\;.
\end{eqnarray}
Since the pressures in the two phases are the same at the transition, 
the bag constant $B$ follows from this requirement, 
\begin{eqnarray}
    P_{\rm EQMD}(\mu_t) - B &=& P_{\rm nucl}(\mu_t)\;.
\end{eqnarray}
The pressure and energy density are then given by
\begin{eqnarray}
    P_{\rm } &=& \begin{cases}
        P_{\rm nucl}\;, &  {\rm if} \quad\mu< \mu_t\;,\\
        P_{\rm EQMD} - B\;, & {\rm if} \quad\mu\geq \mu_t\;, 
    \end{cases}
\end{eqnarray}
and
\begin{eqnarray}
    \epsilon_{\rm } &=& \begin{cases}
        \epsilon_{\rm nucl}\;, &  {\rm if} \quad\mu< \mu_t\;,\\
        \epsilon_{\rm EQMD} + B \;, & {\rm if} \quad\mu\geq \mu_t\;.
    \end{cases}
\end{eqnarray}
The requirement in Eq.~(\ref{conti}) implies that the baryon density is
continuous throughout the density range, giving rise to a smooth pressure
as a function of $\mu$ and 
a continuous quark-hadron transition. We can of course
relax this constraint and consider first-order transitions.

There are two requirements on the equation of state in the entire density range, namely
\begin{enumerate}
    \item The EoS is causal \(0 \leq c_s^2 < 1\).
    \vspace{2mm}
    \item The EoS is stable against density fluctuations.
\end{enumerate}
The first constraint translates into $c_s^{-2}={\mu\over n}{\partial n\over\partial\mu}\geq1$, which implies a minimal slope on the number
density in the $\mu$--$n$ plane. The second constraint implies that
\(-\frac{\partial^2\Omega}{\partial\mu^2} \geq 0\) and therefore that the
thermodynamic potential is a concave function of $\mu$.
We have explicitly checked that our equations of state satisfy the
constraints.

We can choose the parameters in the EQMD model in such a way that the resulting EoS is stiff at intermediate densities and soft at large densities. A large quark condensate, particularly a large strange quark condensate \(\phi_s\) leads to a stiff EoS. A large 
$\omega_{l,s}$ condensate leads to large quark condensates, while a large diquark condensate and large chemical potentials lead to smaller quark condensates. Balancing these mechanisms, we can choose parameters such that we have a stiff EoS at intermediate densities that quickly becomes softer at larger densities.


\section{Numerical results and discussion}
As mentioned above, the parameters in the mesonic sector are determined by matching predictions of the model to physical observables. Ignoring the
breaking of the $U(1)_A$ symmetry due to instantons, we have six parameters in the Lagrangian. We therefore need six observables and we choose the pion decay constant $f_{\pi}$, the light and heavy quark masses $m_l$ and $m_s$, and the meson masses $m_{\pi}$, $m_{\sigma}$, and $m_{\eta^{\prime}}$.
Since $m_{\sigma}$ is a rather broad resonance, we choose a representative value of $m_{\sigma}=500$ MeV. The masses and decay constant are 
\begin{align}
     m_{u,d} &= 300 {\rm MeV}\;, & f_\pi &= 93 {\rm MeV}\;, & m_\pi &= 140 {\rm MeV}\;,\\
    m_\sigma &= 500 {\rm MeV}\;, & m_{\eta'} &= 948 {\rm MeV}\;, & m_\omega &= 783 {\rm MeV}\;.
\end{align}
We find that the results are mainly sensitive to the quark-vector coupling $g_{\omega}$.
We have chosen three parameter sets, where only $g_{\omega}$
is varied from a higher value in set 1 to a lower value in set 3. The values of the parameters are
\begin{align}
    \lambda_3 &= 700\;, & \lambda_4 &= -200\;, & \lambda_5 &= -100\;,\\
    \lambda_\Delta^1\;, &= 100 & \lambda_\Delta^2 &= 100\;, &  g_\Delta &= 0.65 g\;,\\
     m_\Delta &= 1100 {\rm MeV}\;, & \lambda_\omega &= 100\;, & g_\omega^{(1)}
     &= g/3\;,\\
     g_\omega^{(2)} &= g/3.3\;, & g_\omega^{(3)} &= g/3.7\;.
\end{align}

\begin{figure}[htb!]
\centering
    \includegraphics[width=\linewidth]{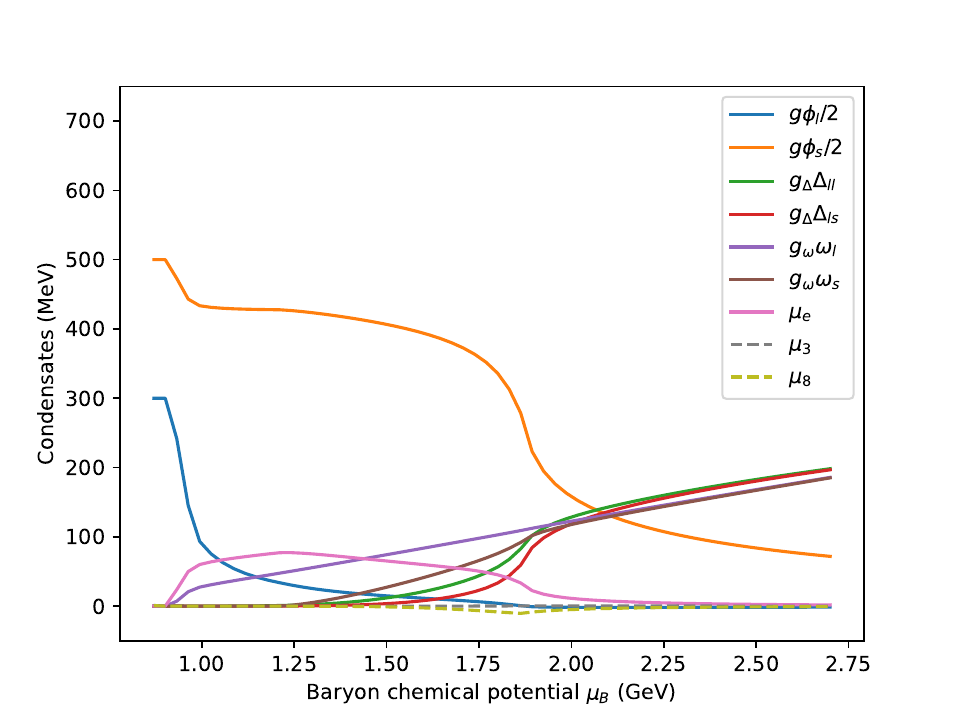}
    \caption{Quark-, diquark- and vector condensates, and chemical potentials as functions of baryon chemical potential for parameter set two. See main text for details.}
    \label{fig:condensates_set2}
\end{figure}

The bag constants are determined by our construction to be 
$B^{1\over4} \approx 234.4$ MeV, $238.2$ MeV, and $241.9$ MeV for parameter sets 1, 2, and 3, respectively. As reported in \cite{baymrev} this is quite close to expectations based on the NJL model, where it is calculated to be \(B^{1\over4}\simeq 218\) MeV or from QCD based calculations where it is estimated to be \(\simeq (250-300)\) MeV\cite{Novikov_NuclPhysB}. By solving the gap equations and neutrality conditions, we obtain the condensates as functions of the baryon chemical potential $\mu_B$ as shown in Fig.~\ref{fig:condensates_set2} for the second parameter set. The quark condensates take on their expected vacuum expectation values. At \(\mu = 300\) MeV, there is a transition from the vacuum phase to the normal quark matter phase. We then find a nonzero expectation value for the omega vector mesons. The diquark gaps are zero in vacuum and in the normal phase become nonzero at around \(\mu_B \approx 1.350\) GeV. This is the transition to the CFL phase. The large difference in the quark condensates, \(\phi_s - \phi_l\), results in the observed difference in the values for the diquark gaps and the gaps of the omega mesons, as well as a nonzero electric chemical potential $\mu_e$. We observe that the differences go to zero as the strange condensate melts. We find that for many of the parameter sets we tested, the two color chemical potentials \(\mu_3\) and \(\mu_8\) play a minimal role, as can be seen in the Fig. \ref{fig:condensates_set2} since they are usually \(\approx 0\).

With the freedom in our parameter space, there is some flexibility in the behavior of the condensates. However, some general behavior is independent of our choice of parameters. The condensates in vacuum are known and specified in the model. At asymptotic densities, we expect to be in the ideal CFL phase. This means that no matter how we choose the parameter sets we will always end up with equal diquark gaps and zero quark condensates at asymptotic densities. It is also possible to show that the $\omega_l$ and $\omega_s$ condensates become equal at higher densities and grow linearly with the chemical potential. How quickly we reach this asymptotic behavior depends on our parameter choices. We will also see later that the speed of sound always goes to the conformal limit in the model.~\footnote{As long as we include a nonzero quartic vector coupling. This requirement was also shown here \cite{Fraga_vectormeson} in a similar context.} For intermediate densities, the existence of a 2SC phase versus a CFL phase, the size of the omega meson condensates and the diquark condensates depend
on the specific choice of parameter values.

\begin{figure}[htb!]
    \centering
    \includegraphics[width=\linewidth]{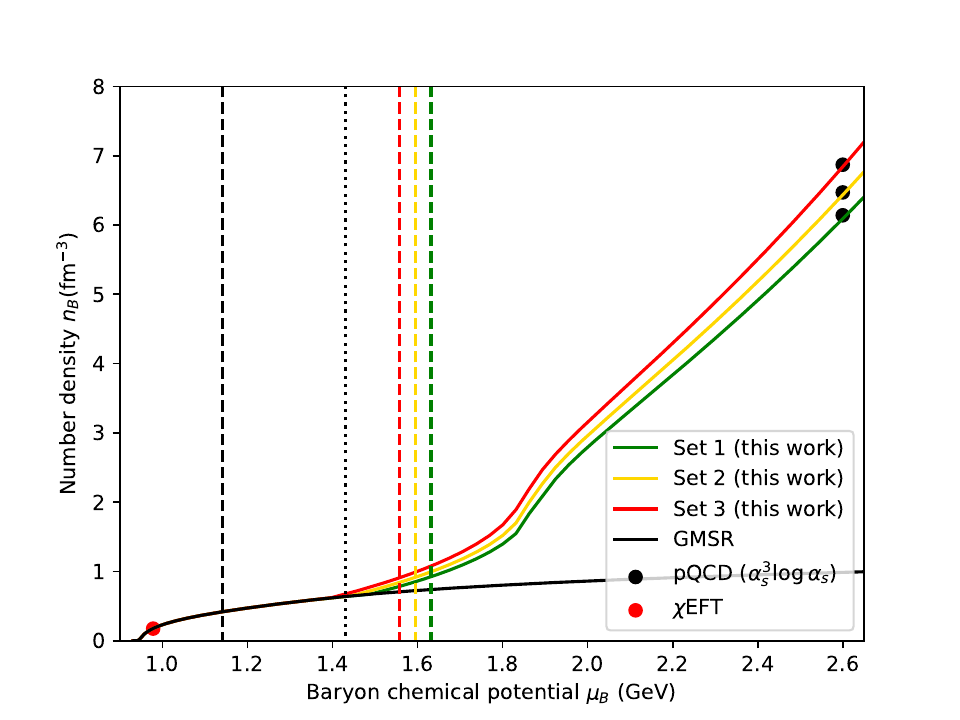}
    \caption{Number density as function of baryon chemical potential for all three parameter sets. The black dots are pQCD for three different values of the renormalization scale and the red dot is $\chi$EFT.
    The black solid line is the density for the nuclear equation of state GMSR~\cite{reddik}. The vertical lines are the values of $\mu_B$ in the center of PSR J0030+0451,
    PSR J0740+6620, and of the maximum mass hybrid stars for the three parameter sets.
    See main text for details.}
    \label{fig:number_density}
\end{figure}

The number density for the three parameter sets is presented in 
Fig.~\ref{fig:number_density}. The red dot is from chiral effective theory, whereas the three black dots are from perturbative QCD
at $\mu=2.6$ GeV. The dots correspond to three different values of the renormalization scale $\Lambda$. 
The vertical lines show the values of the baryon chemical potential in the center of PSR J0030+0451 (dashed line), PSR J0740+6620 (dotted line), and of the maximum mass hybrid stars for the the three parameter sets.
Their central densities are $\approx$
$2.5n_{\rm sat}$, $3.9n_{\rm sat}$, $5.6n_{\rm sat}$, $5.5n_{\rm sat}$, and $5.5n_{\rm sat}$.
Finally, we show the number density for the nuclear EoS in the entire range. We observe in Fig. \ref{fig:number_density} that a larger (smaller) \(g_\omega\) leads to a more slowly (rapidly) increasing density. This results in a stiffer (softer) EoS. 

In the following, we will confront our model with a number of astrophysical
observations
(the errors quoted give one-sigma statistical uncertainties). First, we consider three of the highest pulsar mass measurements, namely:
\begin{enumerate}
\item $M=1.97_{-0.04}^{+0.04}M_{\odot}$ for PSR J1614-2230~\cite{Demorest_nature}. This is a ``Shapiro delay" measurement which uses radio timing observations to detect and model this general-relativistic effect (only possible when the orbit is seen close to edge-on).

\item $M= 2.01_{-0.04}^{+0.04}M_{\odot}$ for PSR J0348+0432~\cite{Antoniadis_Science}. This measurement combines radio-timing and optical spectroscopy with modeling of the white-dwarf companion. It has been recently revised~\cite{Saffer25}, under the assumption that the orbital period derivative is driven exclusively by gravitational-wave emission.

\item $M = 2.27_{-0.15}^{+0.17}M_{\odot}$ for PSR J2215+5135~\cite{Linares_ApJ}. This measurement combines radio-timing and optical spectroscopy with modeling of the strongly irradiated non-degenerate low-mass companion star (see also \cite{Kandel20,Voisin20}).
\vspace{2mm}
\item Spectrophotometry and radio timing observations: $M=2.35_{-0.17}^{+0.17}M_{\odot}$ for PSR J0952-0607 \cite{Romani_ApJ}. This measurement combines radio-timing and optical spectroscopy with modeling of the strongly irradiated non-degenerate ultra-light companion star.
\vspace{2mm}

We also consider simultaneous mass and radius measurements from the NICER collaboration, where two groups have independently reported consistent values \cite{Chat25}, namely:
\vspace{2mm}
    \item $M=1.34^{+0.15}_{-0.16}M_{\odot}$ and $R=12.71_{-1.19}^{+1.14}$ km  for PSR J0030+0451 \cite{Riley_ApJ}. This measurement uses NICER data accumulated over 2 years to model the X-ray pulse profile, with no a-priori knowledge of the neutron star mass.

\item $M=2.073^{+0.069}_{-0.069}M_{\odot}$ and $R=12.49^{+1.28}_{-0.88}$ km for PSR J0740+6620 \cite{Salmi_ApJ}. This measurement combines NICER and XMM data to model the X-ray pulse profile, with a tight prior on the relatively high mass \cite{Fonseca21}.
\end{enumerate}

\begin{figure}[htb!]
    \centering
    \includegraphics[width=\linewidth]{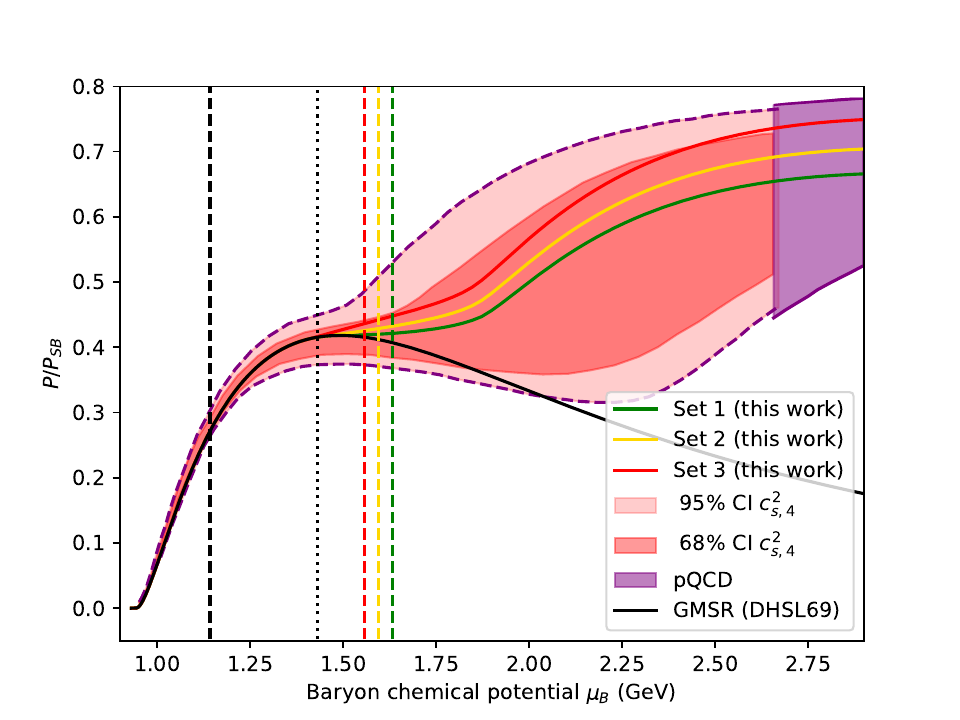}
    \caption{Normalized pressure as a function of $\mu_B$ for three parameter sets. The purple band is pQCD. The black solid line shows the nuclear EoS~\cite{reddik}.  
 The vertical lines have the same meaning as in Fig.~\ref{fig:number_density}.  
    Figure adapted from~\cite{nature}. See main text for details.}
    \label{fig:pressure}
\end{figure}

As mentioned before, we need a realistic nuclear EoS and
there is a plethora of choices~\cite{compose}. We have been guided
by Fig.~\ref{fig:pressure}. We simply require that the pressure
as a function of $\mu_B$ is inside the band. 
We have chosen to use the GMSR EoS~\cite{reddik} from CompOSE library~\cite{compose}. Many of the equations of state either overshoot or undershoot for low densities. The general behavior at intermediate densities for most EoS found in CompOSE is reminiscent of the behavior for the GMSR EoS shown in Fig.~\ref{fig:pressure}, as
shown in Fig. 8 (b) of~\cite{nature}.

The normalized pressure for the three parameter sets
as a function of the baryon chemical potential 
$\mu_B$ is shown in Fig.~\ref{fig:pressure}. The purple band in the figure is the pressure in pQCD to N$^3$LO at \(\mu_B = 2.6\) GeV with values reported in~\cite{Komoltsev_pQCD1}. The band is obtained by varying the renormalization scale
$\Lambda$. 
We also show the 68\% and 95\% credible intervals (CI) from a Bayesian framework presented in \cite{nature}. These intervals were obtained by taking a total of 12 mass-radius measurements of the most massive pulsars, as well as tidal-deformability constraints from the NS-NS merger GW170817 \cite{gw}. 
The vertical lines are the same as in Fig.~\ref{fig:number_density}.
We observe good agreement for a large section of the intermediate chemical potential above \(\mu_B\approx1.5\) GeV. 
We see that the pressure calculated in the EQMD model is consistent with the pQCD constraints, meaning that the total pressure and number density are consistent with both $\chi$EFT and pQCD at the same time. The baryon chemical potential in the center of PSR J0740+6620 is very close to the maximum normalized pressure of the nuclear EoS, which is where we no longer trust it.

\begin{figure}[htb!]
    \centering
    \includegraphics[width=\linewidth]{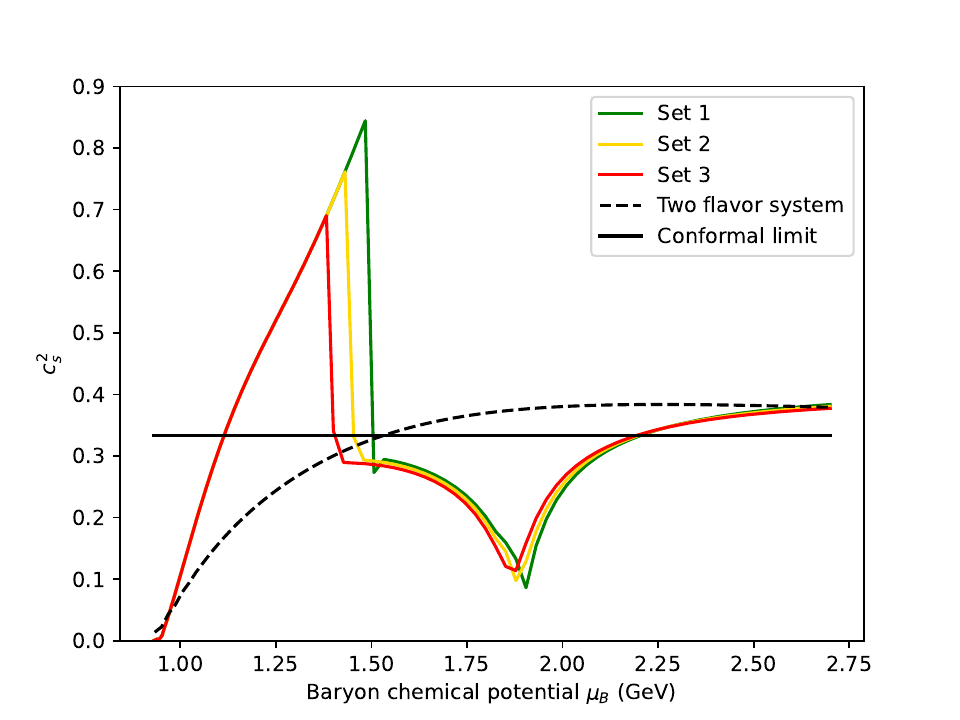}
    \caption{Speed of sound squared $c_s^2$ as a function of baryon chemical potential $\mu_B$ for three different parameter sets. The solid horizontal line is the conformal limit $c_s^2={1\over3}$ and the dashed line is a two-flavor EQMD example without a hybrid construction. See main text for details.}
    \label{fig:cs2}
\end{figure}

The speed of sound for the hybrid star based on the three parameter sets is shown in Fig.~\ref{fig:cs2}. The initial steep increase comes from the GMSR EoS while the sharp drop at $\mu_B \approx 1.5$ GeV reflects the transition to the EQMD EoS. The further sharp decrease in the speed of sound to $c_s^2\approx 0.1$ originates from the rapid softening of the EoS by the melting of the strange quark condensate. For larger chemical potentials, the speed of sound increases above the conformal limit. For very large values of the chemical potential, $c_s^2$ can be shown
to behave as
\begin{eqnarray}
    c_s^{2} \simeq \frac{1}{3}\left[1 + \frac{2}{3}\frac{m_\omega^2\pi^2\gamma_\omega + 2g_\omega g_\Delta^2\Delta^2}{\pi^2\gamma_\omega^3\lambda_\omega\mu^2}\right]\;,\label{eqn:cs2asymptotic} 
\end{eqnarray}
where \(\gamma_\omega = (g_\omega/\lambda_\omega)^{1/3}[\pi^{2/3} + g_\omega(g_\omega/\lambda_\omega)^{1/3}]^{-1}\).
Thus, the speed of sound approaches the conformal limit from above, but the second peak appears for larger values of $\mu_B$ than shown in Figure \ref{fig:cs2}. This limiting behavior is generic for the (E)QM(D) model. The relation in Eq. \eqref{eqn:cs2asymptotic} is only valid for nonzero \(\lambda_\omega\). If we set \(g_\omega\) to zero, it can be shown that we recover the old result of the QMD model derived in~\cite{us3}. This 
result is in contrast to most traditional implementations of the NJL model. For any nonzero vector coupling
$\eta_V$, the speed of sound $c_s$
approaches unity at very high densities~\cite{njl}. This behavior has been remedied in more modern uses of the NJL model by introducing, for example, higher order vector interactions. Moreover, the approach to the conformal limit disagrees with weak-coupling expansions. 
Although the uncertainty bands obtained by varying the renormalization scale are very large, the calculations in~\cite{geisel2,geisel1,minato} suggest that the limit is approached from below.~\footnote{Interestingly, at finite isospin $\mu_I$, the approach to the conformal limit in the quark-meson model~\cite{us,kojo,brandt}
and the NJL model~\cite{ayala} is also from above, but now agrees with weak-coupling calculations~\cite{minato} and lattice simulations~\cite{abbott}.}
Finally, note that the double peak-structure is specific to
three flavors, since it is associated with the melting of the strange quark condensate. 
The density at the minimum of the speed of sound is \(n_B \simeq 1.75 {\rm fm}^{-3}\simeq11n_{\rm sat}\).
This dip in the speed of sound corresponds to the rapid increase 
in the number density in Fig.~\ref{fig:number_density} around \(\mu_B \simeq 1.85\) GeV. The speed of sound for two flavors depends on the
parameters as well, but the shape of the curve is generic.
In order to emphasize the importance of the strange quark, we show a typical two-flavor result (black dashed line).

\begin{figure}[htb!]
    \centering
    \includegraphics[width=\linewidth]{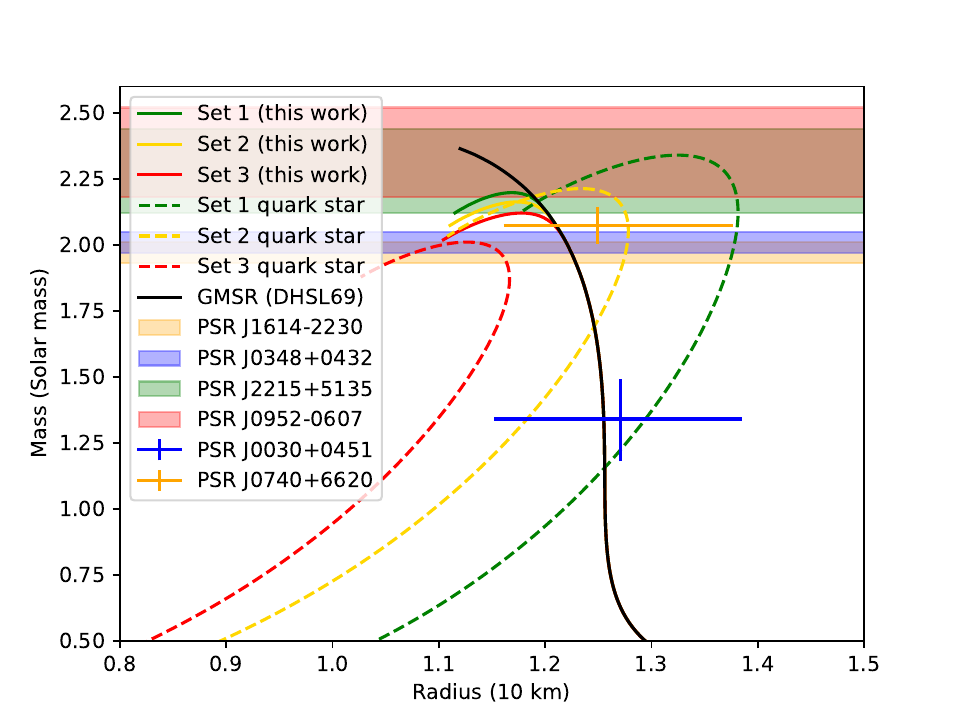}
    \caption{Mass-radius relationship for three different parameter sets. The black line is the nuclear equation of state GMSR~\cite{reddik}
   and the dashes lines are pure quark stars with the same parameters. See main text for details.}
    \label{fig:MassRadius}
\end{figure}

Using the TOV equation with our hybrid EoS as input, we calculate the $M$--$R$ relation shown in Fig.~\ref{fig:MassRadius}. Of the three parameter sets, set one fits with all observations whereas set two and three fit with all observations, except the most massive PSR J0952-0607. A quark core emerges at $M\sim2M_{\odot}$
with $n\approx 3.9 n_{\rm sat}$,
suggesting that PSR J0740+6620 is a borderline case. 
In the three parameter sets, we find that the radius of the quark core
in the maximal mass stars is
$R_1^c \approx 3.8$ km, $R_2^c\approx 4.2$ km, and $R_3^c \approx 4.6$ km. Perhaps counterintuitively, it is the star with the smallest maximum mass (red solid line) that has the largest
quark core. However, for this parameter set, the hybrid-star
$M$--$R$ curve branches off the nuclear $M$--$R$ curve earlier.
We also show pure quark star solutions (dashed lines) for the same parameter sets. 
The idea is that the maximum mass of a hybrid star is determined mainly
by the quark-matter EoS~\cite{GholamiAstro}. A low maximum mass for 
a pure quark star will hardly lead to a very massive hybrid star.

We close this section with a few remarks on future directions.
So far, we have worked exclusively at $T=0$.
The thermodynamic potential in the mean-field approximation can be
straightforwardly generalized to finite temperature. 
This opens up a number of applications. For example, the 
phase diagram in the $\mu$--$T$ plane is of interest in its own right, since there is the possibility of different melting patterns of the diquark condensates, which give rise to new phases. In the context of neutron-star mergers, it is of interest to calculate the mean-free paths of neutrinos for relevant temperatures and densities~\cite{ft1}. Finally, the bulk viscosity can be calculated as a function of temperature and is relevant for the $r$-modes of oscillating compact stars~\cite{ft2}.
\indent
The Lagrangian of the EQMD model has a rather large number of parameters.
The ones belonging to the mesonic sector are determined by matching.
We calculate some physical quantities in the model, such as pole masses and decay constants, and tune the parameters by requiring the model
to reproduce the observed values of these quantities. The remaining parameters are not determined. This leaves us with a large parameter space.
We have, without excessive effort, been able to find parameter sets that
yield hybrid stars whose masses are in agreement with recent astrophysical
observations. This knowledge could serve as a starting point
for a Bayesian analysis. 
In a Bayesian inference, one needs the
prior distributions of the parameters in the Lagrangian as input.
A common choice is a uniform distribution for the parameters, and the parameter sets in this paper give reasonable values for 
the central values and the widths of the distributions.
In this letter, we have focused on the masses and radii of neutron stars, however, other measurements should be included in such an analysis.
For example, the LIGO/VIRGO collaboration provides two-dimensional
posterior probabilities $P(\text{GW}|\tilde{\Lambda},q)$ for the event GW170817, where $\tilde{\Lambda}$ is the tidal deformability and $q$ is the mass ratio of the stars~\cite{gw}.

\section*{Data availability}
Parameter values, data files, and computer code are available at~\cite{github}

\section*{Declaration of competing interest}
The authors declare that they have no known competing ﬁnancial interests
or personal relationships that could have appeared to inﬂuence the
work reported in this paper.

\section*{Acknowledgements}
JOA and MPN thank O. Komoltsev and A. Kurkela for useful discussions of pQCD constraints. ML has received funding from the European Research Council (ERC) under the European Union’s Horizon 2020 research and innovation programme (grant agreement No. 101002352)



\bibliographystyle{elsarticle-harv} 


\end{document}